\documentclass{ieeeaccess}
\usepackage{cite}
\usepackage{amsmath,amssymb,amsfonts}
\usepackage{algorithmic}
\usepackage{graphicx}
\usepackage{textcomp}
\usepackage{academicons}
\usepackage{xcolor}
\usepackage{multirow}
\usepackage{multicol}
\usepackage{makecell}
\usepackage{hyperref}
\hypersetup{
    hidelinks,            
    hypertexnames=false,
}

\definecolor{orcidlogocol}{HTML}{A6CE39}

\usepackage{bm}
\makeatletter
\AtBeginDocument{\DeclareMathVersion{bold}
\SetSymbolFont{operators}{bold}{T1}{times}{b}{n}
\SetSymbolFont{NewLetters}{bold}{T1}{times}{b}{it}
\SetMathAlphabet{\mathrm}{bold}{T1}{times}{b}{n}
\SetMathAlphabet{\mathit}{bold}{T1}{times}{b}{it}
\SetMathAlphabet{\mathbf}{bold}{T1}{times}{b}{n}
\SetMathAlphabet{\mathtt}{bold}{OT1}{pcr}{b}{n}
\SetSymbolFont{symbols}{bold}{OMS}{cmsy}{b}{n}
\renewcommand\boldmath{\@nomath\boldmath\mathversion{bold}}}
\makeatother

\def\BibTeX{{\rm B\kern-.05em{\sc i\kern-.025em b}\kern-.08em
    T\kern-.1667em\lower.7ex\hbox{E}\kern-.125emX}}
    
\begin{document}
\history{Date of publication xxxx 00, 0000, date of current version xxxx 00, 0000.}
\doi{10.1109/ACCESS.2024.0429000}

\title{Advanced POD-Based Performance Evaluation of Classifiers Applied to Human Driver Lane Changing Prediction}
\author{
\uppercase{zahra rastin\textsuperscript{\href{https://orcid.org/0000-0001-9018-3960}{\includegraphics{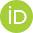}}}} AND
\uppercase{Dirk Söffker\textsuperscript{\href{https://orcid.org/0000-0001-8299-101X}{\includegraphics{orcid.png}}}}, 
\IEEEmembership{Member, IEEE}
}

\address{Chair of Dynamics and Control, University of Duisburg-Essen, 47057 Duisburg, Germany (e-mail: soeffker@uni-due.de)}

\markboth
{Rastin \headeretal: POD-Based Performance Evaluation of Classifiers}
{Rastin \headeretal: POD-Based Performance Evaluation of Classifiers}

\corresp{Corresponding author: Zahra Rastin (e-mail: zahra.rastin@uni-due.de).}

\begin{abstract}
Machine learning (ML) classifiers serve as essential tools facilitating classification and prediction across various domains. The performance of these algorithms should be known to ensure their reliable application. In certain fields, receiver operating characteristic and precision-recall curves are frequently employed to assess machine learning algorithms without accounting for the impact of process parameters. However, it may be essential to evaluate the performance of these algorithms in relation to such parameters. As a performance evaluation metric capable of considering the effects of process parameters, this paper uses a modified probability of detection (POD) approach to assess the reliability of ML-based algorithms. As an example, the POD-based approach is employed to assess ML models used for predicting the lane changing behavior of a vehicle driver. The time remaining to the predicted (and therefore unknown) lane changing event is considered as process parameter. The hit/miss approach to POD is taken here and modified by considering the probability of lane changing derived from ML algorithms at each time step, and obtaining the final result of the analysis accordingly. This improves the reliability of results compared to the standard hit/miss approach, which considers the outcome of the classifiers as either
0 or 1, while also simplifying evaluation compared to the \^{a} versus a approach. Performance evaluation results of the proposed approach are compared with those obtained with the standard hit/miss approach and a pre-developed \^{a} versus a approach to validate the effectiveness of the proposed method. The comparison shows that this method provides an averaging conservative behavior with the advantage of enhancing the reliability of the hit/miss approach to POD while retaining its simplicity.

\end{abstract}

\begin{keywords}
Classification, machine learning, performance evaluation, probability
of detection
\end{keywords}

\titlepgskip=-21pt

\maketitle

\section{Introduction and motivation}
\label{sec:introduction}
\PARstart{S}{upervised} classification is a machine learning (ML) approach that involves training a model using a dataset with known labels, aiming to predict the class or category of new data \cite{RAY2021100011}. Machine learning classifiers are a fundamental part of a wide range of artificial intelligence applications. Object detection, fraud and fault detection, text classification, medical diagnosis, and structural damage detection are examples of tasks in which ML classifiers are extensively employed. 

It is essential to evaluate the performance of ML classifiers as it ensures their reliability and applicability to specific tasks they have been designed for, guiding model selection and improvement. Various evaluation metrics have been used for this purpose. According to \cite{10.1007/978-3-031-35314-7_2}, accuracy, recall, F-score, and precision are the most commonly used ones in recent years \cite{fan2024sampling}, \cite{chen2023classification}, \cite{lalwani2022customer}, \cite{PODDER2021175}, \cite{9121114}. Receiver operating characteristic (ROC) and precision-recall (PR) curves are other frequently employed evaluation tools \cite{AMEYAW2022100220}, \cite{9740533}, \cite{MATINMALAKOUTI2023104796}. The ROC curve is a graphical tool for illustrating a classifier's performance by plotting detection rate (DR) against false alarm rate (FAR) values at various classification thresholds. The area under the ROC curve serves as a summery measure of the model's performance which ranges from 0.5, representing a model with no discriminative power, to 1, denoting a flawless model. Receiver operating characteristic curves can be deceptive when dealing with unbalanced data, making the model's performance seem better than it actually is. To address this common issue in the field of ML, PR  curves are often considered a suitable alternative \cite{10.1145/1143844.1143874}, \cite{10.1371/journal.pone.0118432}, \cite{GIGLIONI2021113029}. On a PR curve, precision is plotted against recall (another name for DR) at various thresholds. Like with ROC curves, a summery measure of a classifier's performance can be obtained by calculating the area under the PR curve which ranges from 0 to 1.

Classification results are often affected by process parameters that are not accounted for by classifiers. Process parameters, which differ from training/model-specific hyperparameters, are task-specific factors that impact recognizability of the target and the final outcome. The size of damage in a damage detection task or the resolution of images used for disease diagnosis tasks are examples of process parameters. Despite their significance, these effects have not been directly accounted in any of the previously mentioned evaluation metrics, therefore have not received sufficient attention in evaluating ML classifiers. To tackle this problem, the probability of detection (POD) approach can be utilized to assess classifiers' performance \cite{ameyaw2019probability}. 

The POD approach is frequently employed for evaluating the efficiency of nondestructive testing (NDT) techniques, and finds applications in safety-critical domains such as aerospace and military fields \cite{doi:10.1177/14759217231193088}, \cite{10.1063/1.4716215}, \cite{hdbk2009nondestructive}. This approach is also attracting attention in other areas where it was less common before, such as structural health monitoring field and nuclear industry \cite{doi:10.1177/14759217211060780}, \cite{s23104813}. Probability of detection-based evaluation results in a curve referred to as the POD curve. In NDT field, the POD curve illustrates the probability of detecting a flaw as a function of its size. This curve can be generated using either binary data  indicating whether the target is detected or not (hit/miss approach), or continuous data providing quantitative assessment of the target (\^{a} versus a approach) \cite{hdbk2009nondestructive}, \cite{virkkunen2019comparison}. 

Ameyaw et al. \cite{9791231} employed the POD approach to assess and compare the performance of ML classifiers. As examples, artificial neural network (ANN), support vector machine (SVM), hidden Markov model (HMM), random forest (RF), and improved versions of these classifiers used for human driver lane changing behavior (LCB) prediction were chosen for performance assessment. As a common driving behavior, lane changing is one of the leading contributors to road accidents and its prediction is essential for autonomous vehicles and  advanced driver-assistance systems \cite{HUANG2024104497}. Modern LCB prediction relies on ML algorithms; however, further improvement in the performance of these algorithms is necessary to facilitate their widespread application in commercial products \cite{XING2020102615}. Ameyaw et al. \cite{9791231} considered the time remaining to the lane changing event as the process parameter. The \^{a} versus a approach to POD was taken using DR values calculated at each time step as the continuous data needed to generate the POD curve. This enabled the comparison and illustration of differences between traditional and enhanced classifiers, in a manner distinct from conventional evaluation metrics such as ROC, DR, and ACC. 

In the previous study \cite{unpublished}, a method for further improving the performance of these algorithms was proposed. Multi-level features extracted from a deep autoencoder were utilized to train an ensemble of classifiers of one type whose hyperparameters were optimized using genetic algorithm. Performance evaluation was applied using the \^{a} versus a approach to POD, employing the lane changing probabilities obtained from ML algorithms at each time point as the continuous response signal, and considering the same process parameter as used in Ameyaw et al. \cite{9791231}. Comparison with results from this paper showed the success of the proposed method in performance enhancement. 

In the present paper, a modified hit/miss approach to POD is taken to assess the classifiers developed in the previous study \cite{unpublished}. In this context, the standard hit/miss approach considers the outcome of the classifier as either 0 (LCB not detected) or 1 (LCB detected). However, in reality, these outcomes are not absolute; rather there is a probability associated with each potential outcome. Therefore, the results from hit/miss approach might not be reliable when comparing the performance of several classifiers to chose the most suitable one for the specified task. To take this fact into account when using hit/miss approach to evaluate classifiers, the final result of the POD analysis is calculated using the probability of lane changing obtained from ML algorithms at each time step. The hit/miss approach is more straightforward than the \^{a} versus a approach. This modification simplifies the performance evaluation process compared to the \^{a} versus a approach proposed in previous contributions \cite{9791231}, \cite{unpublished}
and increases reliability compared to the standard hit/miss approach. The results obtained from the proposed method are compared to those obtained from the standard hit/miss and the \^{a} versus a approaches to POD, validating the effectiveness of the proposed method.

The paper is organized as follows: following the introductory section, an overview of the POD approach and the employed algorithms is provided in Section \ref{sec:background}; the proposed performance evaluation methodology is described in Section \ref{sec:methodology}; in Section \ref{sec:application}, the results of employing the proposed approach is presented and compared to results from standard POD approaches; finally, conclusions are drawn in Section \ref{sec:Conclusion}. 

\section{theoretical background}
\label{sec:background}
\begin{figure}[ht]
    \centering
    \includegraphics[width=\linewidth]{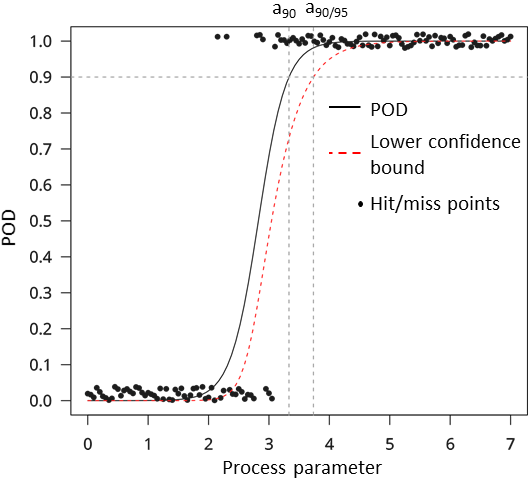}
    \caption{Example of a POD curve obtained from hit/miss data, its 95~\% lower confidence bound, and corresponding $a_{90}$ and $a_{90/95}$ values.}
    \label{fig:example pod}
\end{figure}

This section provides a concise overview of the hit/miss approach to POD and the ML classifiers considered in this paper. The hit/miss approach can be employed to conclude the relationship between detection probability and a considered process parameter using binary (hit/miss) response data. To reach this purpose, a linear model is required to describe the correlation between the continuous process parameter and the binary (0 or 1) data. Ordinary linear regression is not suitable for modeling such a correlation, since it assumes that the response data is continuous and has no bounds. Generalized linear models are used to tackle this problem \cite{hdbk2009nondestructive}. A generalized linear model with a single process parameter can be written in the form of
\begin{equation}g(y)=b_0+b_1a,\label{eq1}\end{equation}
where $b_0$ is the intercept, $b_1$ is the slope, $a$ is the process parameter, and $g(y)$ is a function of the binary response, $y$, that can be used to link $a$ to $y$ through the probability of positive response that continuously changes from 0 to 1 \cite{hdbk2009nondestructive}, \cite{ASTM}. The logistic function is often chosen as the link function, but using the probit function for this purpose is also common. These functions are defined as 
\begin{equation}g(y)=log(\frac{p}{1-p})\label{eq2}\end{equation}
and
\begin{equation}g(y)=\phi^{-1}(p)\label{eq3}\end{equation}
respectively, where $\phi$ is the cumulative standard normal distribution function. Using these functions, the average POD for different process parameter values can be obtained as \cite{ANNIS201398}, \cite{knopp2012considerations}
\begin{equation}POD(a)=\frac{exp(b_0+b_1a)}{1+exp(b_0+b_1a)}\label{eq4}\end{equation}
or
\begin{equation}POD(a)=\phi(b_0+b_1a).\label{eq5}\end{equation}
Depending on the data being modeled, replacing $a$ with $log(a)$ in equations \eqref{eq1}, \eqref{eq4}, and \eqref{eq5} might be more beneficial (further explanations on choosing between $a$ and $log(a)$ are given in Section \ref{sec:methodology}). To take the uncertainties in estimating the parameters of the linear model into account, the likelihood ratio method is utilized to generate the 95 percentile POD curve under which 95 \% of the average POD curves would fall if the study was conducted many times. Process parameter values for 90 \% POD on average and with 95 \% confidence are referred to as $a_{90}$ and $a_{90/95}$ respectively. The $a_{90/95}$ value is frequently used in NDT field to evaluate the effectiveness of defect detection techniques. An example of an average POD curve, its 95 \% lower confidence bound, and corresponding $a_{90}$ and $a_{90/95}$ values, together with the related hit/miss points, is shown in Fig. \ref{fig:example pod}.

Analyzing hit/miss noise is part of the evaluation process that allows considering the false alarm probability (FAP). This probability with 50 \% confidence can be calculated as 
\begin{equation}FAP=\{1+\frac{n-x}{(x+1)F_{(0.5,2x+2,2n-2x)}}\}^{-1},\label{eq6}\end{equation}
where $n$ denotes the number of opportunities for a false alarm, $x$ the number of false alarms, and $F_{(0.5,2x+2,2n-2x)}$ the F-statistics with $(2x+2,2n-2x)$ degrees of freedom and 50~\% (0.5) confidence level \cite{ASTM}.

The described POD approach is used here to evaluate ANN, SVM, HMM, and RF classifiers used for human driver LCB prediction as examples. In contrast to the widely known ANN and SVM models, HMMs are designed to identify dependencies among data points over time and are particularly suited for handling sequential data. Hidden Markov models are statistical frameworks that analyze a series of observable data (here: driving data) emitted by underlying hidden states (here: LCBs). The model subsequently employs inference algorithms to obtain the likelihood of each hidden state at every point along the series of observed data \cite{mor2021systematic}. 

Random forest is another powerful technique that offers robust performance for a variety of classification and regression tasks. This algorithm combines the outputs of several decision trees to develop a more robust model. A decision tree is a hierarchical model that splits data into branches based on features that best separate them, leading to final outcomes at leaf nodes. Each tree is trained on a different portion of training data, which ensures diversity among the trees and reduces overfitting. For classification, RFs aggregate the predictions of individual trees through a majority voting process. 

\section{modified hit/miss approach to POD}
\label{sec:methodology}
The methodology for modifying the hit/miss approach to POD to evaluate ML classifiers is explained in this section. This includes taking the probability of hit/miss data obtained from ML algorithms for different process parameter values into account for performance evaluation. The suggested methodology is illustrated in Fig. \ref{fig:hit miss}. First, test data corresponding to various process parameter values are fed into a pre-trained ML model. For instance, in the LCB prediction task, where the process parameter is the time remaining until the lane changing event, the trained classifier is provided with driving data recorded over time. It is assumed that the probability of the classifier detecting the target for a given process parameter value is determined based on the results of 10 separate experiments conducted with that specific process parameter value. The number of experiments that result in target detection $(n)$ is calculated by multiplying this probability (P) by 10 and rounding the result
\begin{equation}n=round(10×P).\label{eq7}\end{equation}
Therefore, considering each process parameter value, $n$ experiments out of 10 lead to target detection (hit/1), and the rest fail to detect the target (miss/0). 

\Figure[t!](topskip=0pt, botskip=0pt, midskip=0pt){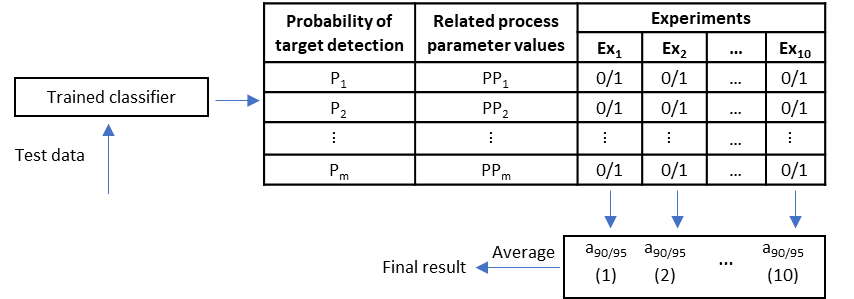}
{ \textbf{Modified hit/miss approach to POD considering m different values of process parameter.}\label{fig:hit miss}}

\Figure[t!](topskip=0pt, botskip=0pt, midskip=0pt){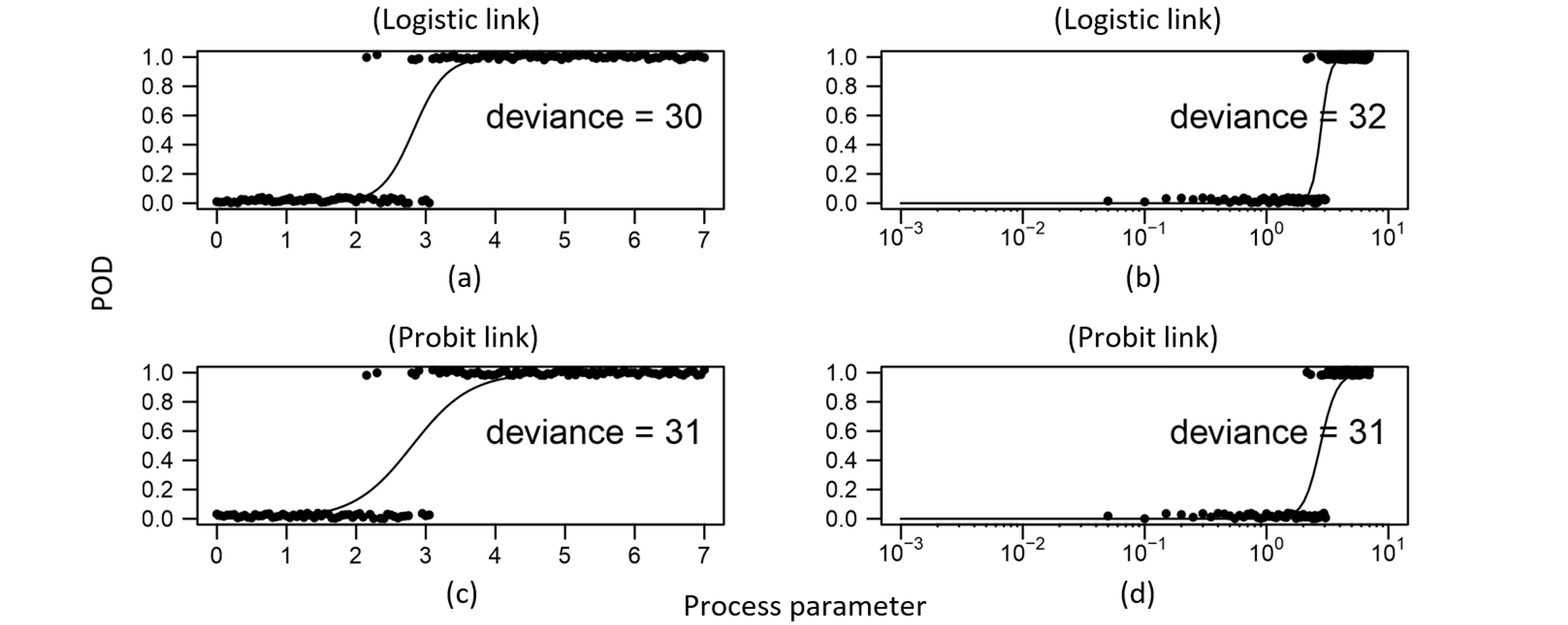}
{\textbf{Four POD models with (a) Cartesian horizontal axis and logistic link, (b) logarithmic horizontal axis and logistic link, (c) Cartesian horizontal axis and probit link, and (d) logarithmic horizontal axis and probit link.}\label{fig:possible models}}

The standard hit/miss approach is applied to 0/1 results from each experiment separately. For each experiment, four POD models are considered, utilizing combinations of logarithmic or Cartesian horizontal axes, with either the logistic or the probit link function, as shown in Fig. \ref{fig:possible models}. Among these models, the one with the lowest deviance is chosen for obtaining the POD curve. The deviance quantifies the overall spread of the data, so lower values are preferable. The $a_{90/95}$ values related to the considered experiments are extracted from the generated POD curves. The average of these values is calculated to obtain the final $a_{90/95}$ for the considered ML classifier. Additionally, the PFA can be calculated using equation \eqref{eq6}.  

\section{machine learning classifiers}
\label{sec:classifiers}
Machine learning classifiers considered for performance evaluation in this study include ANNs, SVMs, HMMs, and RFs developed previously for human driver LCB prediction. The methodology for LCB prediction using these classifiers, and employing the \^{a} versus a approach to POD to assess them is thoroughly described in \cite{unpublished} and is repeated here briefly.

In the first step of the proposed methodology, the features required as inputs to ML models were extracted using a deep autoencoder. Driving data regarding the status of the considered vehicle and its surrounding vehicles were used to train the autoencoder. After training, the encoder part of the network, including 4 hidden layers with 24, 16, 8, and 4 neurons, was utilized to extract multi-level features from the data, which were then fed to ML classifiers.  

Three classes of LCB were considered: lane changing to left (LCL), lane changing to right (LCR), and lane keeping. Features from each encoder layer were utilized to train two binary classifiers of each type, where each classifier regarded either LCL or LCR as the positive class and remaining two classes as negative. This process was done considering features from 4 encoder layers, 4 classifier types, and two binary classifiers of each type, resulting in a total of 32 trained models. The hyperparameters of all ML models were optimized using a genetic algorithm. 

Considering the time remaining until the lane changing event as the process parameter, and the probability of detecting LCBs at each time point as the response signal, the \^{a} versus a approach to POD was employed to assess the classifiers using test data, and specify the algorithms that were able to detect LCBs earlier. The winner-take-all ensemble strategy was applied to $a_{90/95}$ values from classifiers of the same type trained to predict the same LCB. Using this strategy, multiple classifiers compete and the one with the best performance determines the ensemble output. In the next section, the 32 classifiers developed in \cite{unpublished} will be evaluated using the modified hit/miss approach introduced in Section \ref{sec:methodology} and the results will be compared to those from standard POD approaches.

\section{APPLICATION AND RESULTS}
\label{sec:application}
\begin{figure}[ht]
    \centering
    \includegraphics[width=\linewidth]{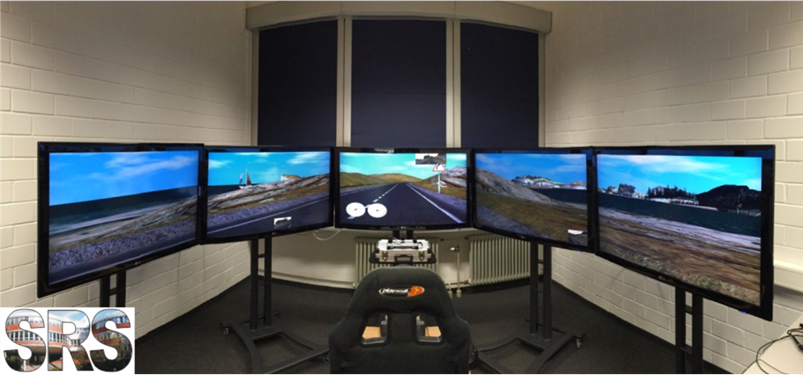}
    \caption{SCANeR™ studio, Chair of Dynamics and Control, University of Duisburg–Essen, Germany.}
    \label{fig:Driving_simulator}
\end{figure}

\begin{figure}[ht]
    \centering
    \includegraphics[width=\linewidth]{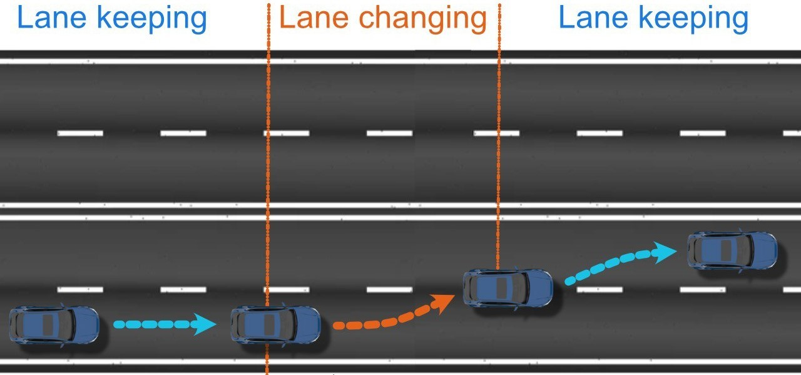}
    \caption{The driving environment and lane changing/keeping
behaviors.}
    \label{fig:LCB}
\end{figure}

The modified and standard hit/miss approaches to POD are used to evaluate the classifiers trained in \cite{unpublished}. Data obtained from the SCANeR™ studio driving simulator shown in Fig. \ref{fig:Driving_simulator} was utilized for this purpose. The simulator uses virtual sensors such as cameras, radar, and lasers to collect data, providing a comprehensive understanding of the vehicle's environment. The simulated driving environment consists of a highway with two lanes in each direction. Three drivers, aged 25 to 38 and holding valid driver's licenses, were recruited to collect the required data. Each participant drove for approximately 40 minutes to obtain the training dataset, with an additional 10 minutes of driving data recorded for the test phase. During the simulation, drivers were permitted to overtake slower vehicles and return to their original lane \cite{8917489}. The vehicle's lane was determined by the position of its center point, and lane changes were identified by shifts in this position. A lane changing event began with the last significant steering wheel adjustment, and the period from this adjustment to the lane shift was defined as the lane changing interval \cite{AMEYAW2022100220}. The driving environment and lane changing/keeping behaviors are depicted in Fig. \ref{fig:LCB}.

Training and test dataset samples, recorded every 0.05 s, comprised  26 variables including the velocity of the considered ego vehicle and surrounding vehicles, distances to surrounding vehicles, time to collision, lane number, turn signal status, gear engaged, steering wheel angle, heading angle, accelerator position, and brake pressure of the ego vehicle \cite{8917489}. The test dataset is utilized here for classifiers' performance evaluation. Categorical variables with no inherent order are one-hot-encoded. The data are then normalized between 0 and 1, and fed to the trained autoencoder \cite{unpublished} for multi-level feature extraction. 

\begin{figure}[ht]
    \centering
    \includegraphics[width=\linewidth]{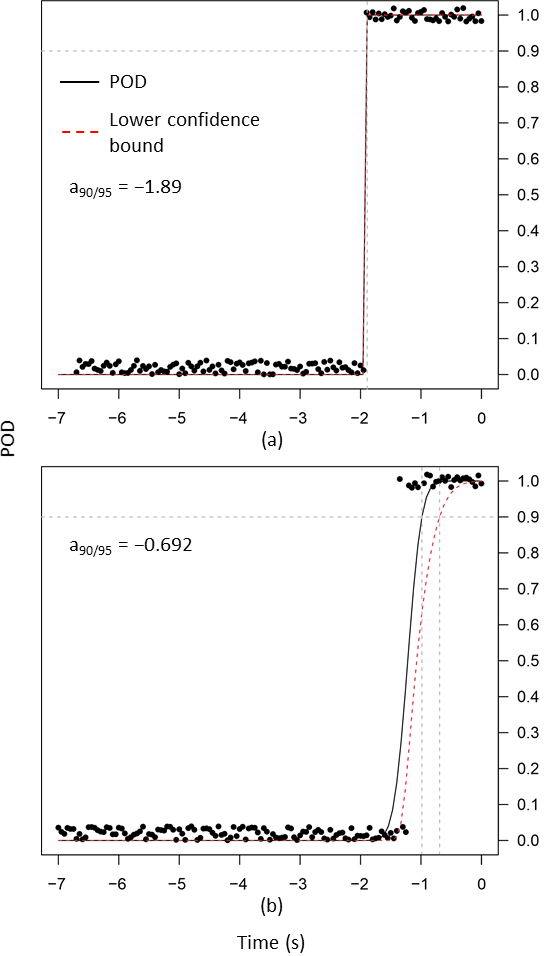}
    \caption{Probability of detection curves obtained from the standard hit/miss approach for (a) the ANN and (b) the SVM trained on features from the first encoder layer to predict LCL using test data from the first driver.}
    \label{fig:example ann svm}
\end{figure}

Features from each encoder layer are input to the 8 related trained classifiers (2 binary ANNs, SVMs, HMMs, and RFs), and the probability that each data sample belongs to the positive class (LCL/LCR class) is extracted from the ML models. The time frames from 7 s before the lane change to the moment of the event are considered, and the average of the extracted probabilities at each time point within this 7-second span is computed. To apply the modified hit/miss approach, it is assumed that the average probability is the result of 10 separate experiments, as explained in Section \ref{sec:methodology}. 
The standard hit/miss approach can also be applied, defining a hit (1) as an average probability greater than 50 \%, and a miss (0) as an average probability of 50 \% or less. Examples of POD curves obtained from the standard hit/miss approach for the ANN and the SVM trained on features from the first encoder layer to predict LCL using test data from the first driver are shown in Fig. \ref{fig:example ann svm}. In this figure, $a_{90/95}=-1.89$ and $a_{90/95}=-0.692$ mean the algorithms can reliably predict LCBs 1.89 s and 0.692 s before the event respectively. Clearly, the earlier the algorithm predicts the LCB, the better its performance.

The $a_{90/95}$ values obtained using the standard and the modified hit/miss approaches to evaluate ANNs and SVMs trained on different encoder layers to predict LCL/LCR, when tested on data from each driver separately, are presented in Tables \ref{tab1} and \ref{tab2}. The results from the \^{a} versus a approach to POD from the previous study \cite{unpublished} are also included in the table, along with the differences between the $a_{90/95}$ values from the \^{a} versus a approach and the two other approaches, to facilitate comparison between these methods. According to these tables, $a_{90/95}$ results from the modified hit/miss approach are generally smaller than those from the standard version. Additionally, they are closer to \^{a} versus a approach results (exceptions are highlighted in red). This outcome is expected, as the \^{a} versus a approach used probabilities of predicting LCBs at different time points as response values in its evaluation; by modifying the hit/miss approach to consider these probabilities instead of simple 0/1 values, the final result is anticipated to align more closely with the result from the \^{a} versus a approach. Similar outcomes were obtained for HMMs and RFs, but they are not included here for the sake of brevity.

\begin{table}
\caption{\textbf{Comparison of $a_{90/95}$ values from different POD approaches applied to ANN and SVM performance evaluation for LCL prediction.}}
\label{table}
\setlength{\tabcolsep}{3pt}
\begin{tabular}{|p{5pt}|p{15pt}|p{5pt}|p{25pt}|p{25pt}|p{25pt}|p{45pt}|p{45pt}|}
\hline
\multirow{2}{*}{\rotatebox{90}{Algorithm}}&
\rule{0pt}{2.5ex} 
\multirow{2}{*}{\rotatebox{90}{\shortstack{Related \\layer}}}&
\multirow{2}{*}{\rotatebox{90}{Driver no.}}&
\multicolumn{3}{c|}{\shortstack{$a_{90/95}$ \\(seconds before the event)}}&
\multicolumn{2}{c|}{Difference in $a_{90/95}$ values (s)}
\\
\cline{4-8}
&
&
&
\^{a} vs a &
SHM &
MHM &
\^{a} vs a and SHM &
\^{a} vs a and MHM\\
\hline
\multirow{12}{*}{\rotatebox{90}{ANN}}&
\multirow{3}{*}{1}&
1 &
3.906 &
1.89 &
2.778 &
2.016 &
1.128\\
\cline{3-8}
&
&
2 &
4.816&
2.03 &
2.734 &
2.786 &
2.082\\
\cline{3-8}
&
&
3 &
1.603 &
0 &
1.005 &
1.603 &
0.598\\
\cline{2-8}
&
\multirow{3}{*}{2}&
1 &
2.367 &
1.89 &
2.840 &
0.477 &
0.473\\
\cline{3-8}
&
&
2 &
4.832 &
2.33 &
3.168 &
2.502 &
1.664\\
\cline{3-8}
&
&
3 &
3.133 &
0 &
1.191 &
3.133 &
1.942\\
\cline{2-8}
&
\multirow{3}{*}{3}&
1 &
5.442 &
4.685 &
4.205 &
0.757 &
\textcolor{red}{1.237}\\
\cline{3-8}
&
&
2&
5.032&
3.57&
3.397&
1.462&
\textcolor{red}{1.634}\\
\cline{3-8}
&
&
3&
3.23&
0&
1.472&
3.23&
1.758\\
\cline{2-8}
&
\multirow{3}{*}{4}&
1 &
5.412 &
5.924 &
4.444 &
0.512 &
\textcolor{red}{0.968}\\
\cline{3-8}
&
&
2&
2.297&
4.27&
3.315&
1.973&
1.018\\
\cline{3-8}
&
&
3&
4.095&
0&
1.632&
4.095&
2.463\\
\hline
\multirow{12}{*}{\rotatebox{90}{SVM}}&
\multirow{3}{*}{1}&
1 &
2.332 &
0.692 &
1.628 &
1.64 &
0.704\\
\cline{3-8}
&
&
2&
4.105&
1.19&
1.874&
2.915&
2.231\\
\cline{3-8}
&
&
3&
2.784&
0&
0.7&
2.784&
2.084\\
\cline{2-8}
&
\multirow{3}{*}{2}&
1 &
2.891 &
1.06 &
2.003 &
1.831 &
0.888\\
\cline{3-8}
&
&
2&
4.649&
1.96&
2.464&
2.686&
2.182\\
\cline{3-8}
&
&
3&
2.124&
0&
0.833&
2.124&
1.291\\
\cline{2-8}
&
\multirow{3}{*}{3}&
1 &
5.073 &
3.443 &
3.928 &
1.63 &
1.145\\
\cline{3-8}
&
&
2&
4.931&
3.57&
3.452&
1.361&
\textcolor{red}{1.479}\\
\cline{3-8}
&
&
3&
4.029&
0&
1.495&
4.029&
2.533\\
\cline{2-8}
&
\multirow{3}{*}{4}&
1 &
5.78 &
3.501 &
3.625 &
2.279 &
2.154\\
\cline{3-8}
&
&
2&
4.474&
1.827&
3.281&
2.647&
1.193\\
\cline{3-8}
&
&
3&
4.021&
0&
1.445&
4.021&
2.576\\
\hline
\end{tabular}
\vskip 2pt
\noindent{\footnotesize{Legend- SHM: standard hit/miss, MHM: modified hit/miss, {\textcolor{red}{red}}: the $a_{90/95}$ value from the \^{a} versus a approach is farther from that of the modified hit/miss approach than from that of the standard version.}}
\label{tab1}
\end{table}

\begin{table}
\caption{\textbf{Comparison of $a_{90/95}$ values from different POD approaches applied to ANN and SVM performance evaluation for LCR prediction.}}
\label{table}
\setlength{\tabcolsep}{3pt}
\begin{tabular}{|p{5pt}|p{15pt}|p{5pt}|p{25pt}|p{25pt}|p{25pt}|p{45pt}|p{45pt}|}
\hline
\multirow{2}{*}{\rotatebox{90}{Algorithm}}&
\rule{0pt}{2.5ex} 
\multirow{2}{*}{\rotatebox{90}{\shortstack{Related \\layer}}}&
\multirow{2}{*}{\rotatebox{90}{Driver no.}}&
\multicolumn{3}{c|}{\shortstack{$a_{90/95}$ \\(seconds before the event)}}&
\multicolumn{2}{c|}{Difference in $a_{90/95}$ values (s)}
\\
\cline{4-8}
&
&
&
\^{a} vs a &
SHM &
MHM &
\^{a} vs a and SHM &
\^{a} vs a and MHM\\
\hline
\multirow{12}{*}{\rotatebox{90}{ANN}}&
\multirow{3}{*}{1}&
1 &
1.878 &
1.19 &
1.666 &
0.688 &
0.212\\
\cline{3-8}
&
&
2 &
2.913 &
0.98 &
1.497 &
1.933 &
1.416\\
\cline{3-8}
&
&
3 &
2.132 &
1.06 &
1.528 &
1.072 &
0.604\\
\cline{2-8}
&
\multirow{3}{*}{2}&
1 &
1.585 &
1.19 &
1.715 &
0.395 &
0.13\\
\cline{3-8}
&
&
2 &
2.914 &
1.05 &
1.627 &
1.864 &
1.287\\
\cline{3-8}
&
&
3 &
1.615 &
1.06 &
1.436 &
0.555 &
0.179\\
\cline{2-8}
&
\multirow{3}{*}{3}&
1 &
4.177 &
0.929 &
1.975 &
3.248 &
2.202\\
\cline{3-8}
&
&
2&
3.789&
1.05&
2.245&
2.739&
1.544\\
\cline{3-8}
&
&
3&
2.248&
1.543&
2.722&
0.705&
0.474\\
\cline{2-8}
&
\multirow{3}{*}{4}&
1 &
2.157 &
1.051 &
1.172 &
1.106 &
0.985\\
\cline{3-8}
&
&
2&
2.275&
0.366&
0.794&
1.909&
1.481\\
\cline{3-8}
&
&
3&
2.137&
0.493&
0.98&
1.644&
1.157\\
\hline
\multirow{12}{*}{\rotatebox{90}{SVM}}&
\multirow{3}{*}{1}&
1 &
2.111 &
1.051 &
1.352 &
1.06 &
0.758\\
\cline{3-8}
&
&
2&
3.639&
1.05&
1.93&
2.589&
1.709\\
\cline{3-8}
&
&
3&
2.262&
0.33&
2.106&
1.932&
0.156\\
\cline{2-8}
&
\multirow{3}{*}{2}&
1 &
1.457 &
0.561 &
0.559 &
0.896 &
\textcolor{red}{0.898}\\
\cline{3-8}
&
&
2&
3.521&
0.98&
1.484&
2.541&
2.037\\
\cline{3-8}
&
&
3&
2.645&
0.561&
1.393&
2.084&
1.252\\
\cline{2-8}
&
\multirow{3}{*}{3}&
1 &
0.084 &
0.98 &
1.112 &
0.896 &
\textcolor{red}{1.028}\\
\cline{3-8}
&
&
2&
3.003&
0.98&
1.704&
2.023&
1.299\\
\cline{3-8}
&
&
3&
2.192&
0&
2.023&
2.192&
0.168\\
\cline{2-8}
&
\multirow{3}{*}{4}&
1 &
1.709 &
0.98 &
1.137 &
0.729 &
0.572\\
\cline{3-8}
&
&
2&
3.581&
1.05&
1.685&
2.531&
1.896\\
\cline{3-8}
&
&
3&
2.282&
0.038&
1.96&
2.244&
0.322\\
\hline
\end{tabular}
\vskip 2pt
\noindent{\footnotesize{{Legend- SHM: standard hit/miss, MHM: modified hit/miss, {\textcolor{red}{red}}: the $a_{90/95}$ value from the \^{a} versus a approach  is farther from that of the modified hit/miss
approach than from that of the standard version.}}}
\label{tab2}
\end{table}

Like in the previous study, the winner-take-all ensemble strategy is applied to performance evaluation results from ML models of the same type having the same task. The final results for each classifier type, when tested on data from each driver and evaluated using the three POD approaches, along with the corresponding encoder layers are detailed in Tables \ref{tab3} and \ref{tab4}. False alarm probabilities calculated using equation \ref{eq6}, considering encoder layers related to the modified hit/miss approach, are also included in the tables. It can be seen that in general, the best encoder layers according to the \^{a} versus a approach are more similar to those identified by the modified hit/miss approach than to those identified by the standard hit/miss approach (18 out of 24 similar cases versus 9 out of 24). Therefore, the modified hit/miss approach is more reliable when deciding about the most suitable set of features for ML classifiers. Furthermore, the best $a_{90/95}$ values for each driver are shown in green in Tables \ref{tab3} and \ref{tab4}. According to all three approaches, ANNs are the most reliable algorithms for LCB prediction.

\begin{table}
\caption{\textbf{Final performance evaluation results according to the three considered approaches to POD for LCL prediction.}}
\label{table}
\setlength{\tabcolsep}{3pt}
\begin{tabular}{|p{5pt}|p{5pt}|p{25pt}|p{25pt}|p{25pt}|p{25pt}|p{25pt}|p{25pt}|p{20pt}|}
\hline
\multirow{2}{*}{\makebox[0pt][c]{\rotatebox{90}{\parbox[c]{0.9cm}{Algorithm}}}}&
\multirow{2}{*}{%
    \parbox[c][\ht\strutbox][c]{2cm}{\rotatebox{90}{Driver no.}}}&
\multicolumn{3}{c|}{\shortstack{$a_{90/95}$ \\(seconds before the event)}}&
\multicolumn{3}{c|}{Related encoder layer}&
\rule{0pt}{5ex}
\multirow{2}{*}{FAP}
\\
\cline{3-8}
\rule{0pt}{5ex}
&
&
\^{a} vs a &
SHM &
MHM &
\^{a} vs a &
SHM &
MHM &
\\
\hline
\multirow{3}{*}{\rotatebox{90}{ANN}}&
1 &
5.442 &
\textcolor{green}{5.924} &
\textcolor{green}{4.443} &
3 &
4&
4&
\multirow{3}{*}{0.062}\\
\cline{2-8}
&
2&
\textcolor{green}{5.032} &
\textcolor{green}{4.27} &
3.397 &
3 &
4 &
3 &
\\
\cline{2-8}
&
3&
\textcolor{green}{4.095} &
0 &
\textcolor{green}{1.632} &
4 &
All &
4 &\\
\hline
\multirow{3}{*}{\rotatebox{90}{SVM}}&
1 &
\textcolor{green}{5.78} &
3.501 &
3.928 &
4 &
4 &
3 &
\multirow{3}{*}{0.039}\\
\cline{2-8}
&
2&
4.931 &
3.57 &
\textcolor{green}{3.452} &
3 &
3 &
3 &
\\
\cline{2-8}
&
3&
4.029 &
0 &
1.496 &
3 &
All &
3 &\\
\hline
\multirow{3}{*}{\rotatebox{90}{HMM}}&
1 &
5.248 &
2.52 &
3.108 &
2 &
2 &
2 &
\multirow{3}{*}{0.052}\\
\cline{2-8}
&
2&
4.299 &
1.89 &
2.772 &
2 &
2 &
2 &
\\
\cline{2-8}
&
3&
3.467 &
\textcolor{green}{0.275} &
1.274 &
2 &
2 &
2 &\\
\hline
\multirow{3}{*}{\rotatebox{90}{RF}}&
1 &
1.298 &
0.071 &
0.329 &
2 &
2\&3 &
2 &
\multirow{3}{*}{0.0001}\\
\cline{2-8}
&
2&
3.529 &
0 &
1.057 &
1 &
All &
1 &
\\
\cline{2-8}
&
3&
4.025 &
0 &
0.486 &
3 &
All &
3 
&\\
\hline
\end{tabular}
\vskip 2pt
\noindent{\footnotesize{Legend- {\textcolor{green}{green}}: the best $a_{90/95}$ value for each driver, according to each POD approach.}}
\label{tab3}
\end{table}

\begin{table}
\caption{\textbf{Final performance evaluation results according to the three considered approaches to POD for LCR prediction.}}
\label{table}
\setlength{\tabcolsep}{3pt}
\begin{tabular}{|p{5pt}|p{5pt}|p{25pt}|p{25pt}|p{25pt}|p{25pt}|p{25pt}|p{25pt}|p{20pt}|}
\hline
\multirow{2}{*}{\makebox[0pt][c]{\rotatebox{90}{\parbox[c]{0.9cm}{Algorithm}}}}&
\multirow{2}{*}{%
    \parbox[c][\ht\strutbox][c]{2cm}{\rotatebox{90}{Driver no.}}}&
\multicolumn{3}{c|}{\shortstack{$a_{90/95}$ \\(seconds before the event)}}&
\multicolumn{3}{c|}{Related encoder layer}&
\rule{0pt}{5ex}
\multirow{2}{*}{FAP}
\\
\cline{3-8}
\rule{0pt}{5ex}
&
&
\^{a} vs a &
SHM &
MHM &
\^{a} vs a &
SHM &
MHM &
\\
\hline
\multirow{3}{*}{\rotatebox{90}{ANN}}&
1 &
\textcolor{green}{4.177} &
\textcolor{green}{1.19} &
\textcolor{green}{1.975} &
3 &
1\&2&
3&
\multirow{3}{*}{0.065}\\
\cline{2-8}
&
2&
\textcolor{green}{3.789} &
\textcolor{green}{1.05} &
\textcolor{green}{2.245} &
3 &
2\&3 &
3&
\\
\cline{2-8}
&
3&
2.248 &
\textcolor{green}{1.543} &
\textcolor{green}{2.722} &
3 &
3 &
3&\\
\hline
\multirow{3}{*}{\rotatebox{90}{SVM}}&
1 &
2.111 &
1.051 &
1.352 &
1 &
1 &
1&
\multirow{3}{*}{0.027}\\
\cline{2-8}
&
2&
3.639 &
\textcolor{green}{1.05} &
1.93 &
1 &
1\&4 &
1&
\\
\cline{2-8}
&
3&
\textcolor{green}{2.645} &
0.561 &
2.106 &
2 &
2 &
1&\\
\hline
\multirow{3}{*}{\rotatebox{90}{HMM}}&
1 &
2.834 &
1.051 &
1.974 &
3 &
3&
3&
\multirow{3}{*}{0.037}\\
\cline{2-8}
&
2&
3.161 &
0.701 &
0.937 &
3 &
2\&3 &
3&
\\
\cline{2-8}
&
3&
2.254 &
0.561 &
0.998 &
2 &
4 &
2&\\
\hline
\multirow{3}{*}{\rotatebox{90}{RF}}&
1 &
1.157 &
0.071 &
0.397 &
4 &
All&
1&
\multirow{3}{*}{0.0027}\\
\cline{2-8}
&
2&
2.555 &
0 &
0.663 &
4 &
All &
1&
\\
\cline{2-8}
&
3&
1.661 &
0.42 &
0.563 &
4 &
2 &
2&\\
\hline
\end{tabular}
\vskip 2pt
\noindent{\footnotesize{Legend- {\textcolor{green}{green}}: the best $a_{90/95}$ value for each driver, according to each POD approach.}}
\label{tab4}
\end{table}

Overall, the results show that modifying the hit/miss approach brings its outcomes closer to those from the \^{a} versus a approach. These results are more reliable than those based solely on 0/1 values when determining the suitable set of features for the considered ML classifiers and assessing the algorithms' reliability in predicting upcoming LCBs early enough. 
Therefore, this modification allows the hit/miss approach to be used for evaluating ML classifiers in a more reliable way, while remaining simpler than the \^{a} versus a approach.

\section{Conclusion}
\label{sec:Conclusion}
This paper proposes POD-based performance evaluation of ML classifiers using a modified hit/miss approach. Unlike common performance evaluation metrics typically used for assessing ML algorithms, the POD-based approach considers the effects of process parameters in evaluation, offering new insights into the reliability of ML algorithms. The hit/miss approach to POD is modified here by incorporating the probability of target detection derived from ML algorithms for various process parameter values and obtaining the final result of the analysis accordingly. This enhances the reliability of the results compared to the standard hit/miss approach, which considers classifier outcomes as either 0 or 1, as in reality, these outcomes are probabilistic rather than absolute.

As an example, the POD-based approach is used to evaluate ML models trained on multi-level features from a deep autoencoder for predicting drivers' LCB using data from a driving simulator. The time until the predicted (and unknown) lane changing event is taken as the process parameter, and the classifiers' ability to predict the intended driver behavior is assessed relative to the time remaining before the event.

Performance evaluation results from the proposed modified hit/miss approach are compared to those from the standard hit/miss approach and the \^{a} versus a approach to POD. Modifying the hit/miss approach makes its outcomes more similar to those from the â versus a approach. These results are more reliable than those based solely on 0/1 values when selecting appropriate features and evaluating the algorithms' effectiveness in predicting upcoming LCBs early. The hit/miss approach is more straightforward than the â versus a approach to POD; this modification enhances its reliability while maintaining its simplicity.

\bibliographystyle{ieeetr}
\bibliography{mybibliography}

\begin{IEEEbiography}[{\includegraphics[width=1in,height=1.25in,clip,keepaspectratio]{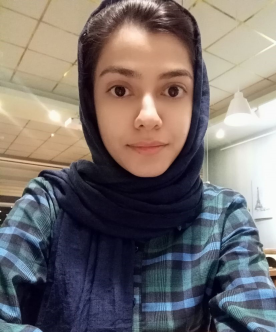}}]{Zahra Rastin} received the M.Sc.
degree in civil engineering from
Iran University of Science and
Technology in 2020. She is currently
pursuing a Ph.D. in mechanical engineering at University of Duisburg-Essen, Germany. Her research interests include structural health monitoring, damage detection, machine learning, and developing reliability evaluation methods for
machine learning approaches.
\end{IEEEbiography}

\begin{IEEEbiography}[{\includegraphics[width=1in,height=1.25in,clip,keepaspectratio]{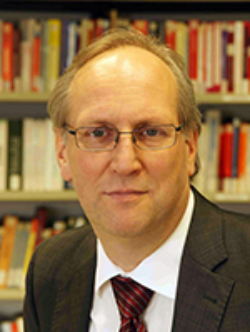}}]{DIRK SÖFFKER} (Member, IEEE) received the
Dr.-Ing. degree in mechanical engineering and
the Habilitation degree in automatic control/safety
engineering from the University of Wuppertal,
Germany, in 1995 and 2001, respectively. Since
2001, he leads the Chair of Dynamics and
Control, University of Duisburg-Essen, Germany.
His current research interests include diagnostics
and prognostics, modern methods of control theory, safe human interaction with technical systems,
safety and reliability control engineering of technical systems, and cognitive
technical systems.
\end{IEEEbiography}

\EOD

\end{document}